\documentclass[12pt,preprint]{aastex}

\slugcomment{}
\shorttitle{Beyond the Coronal Graveyard}
\shortauthors{T.\ R.\ Ayres}
\begin{document}

\title{Beyond the Coronal Graveyard}

\author{Thomas R.\  Ayres}

\affil{Center for Astrophysics and Space Astronomy,\\
389~UCB, University of Colorado,
Boulder, CO 80309;\\ Thomas.Ayres@Colorado.edu}

\begin{abstract}
New {\em Chandra}\/ High Resolution Camera pointings on the ``non-coronal'' red giant Arcturus (HD\,124897; $\alpha$~Boo: K1.5~III) corroborate a tentative soft X-ray detection in a shorter exploratory exposure sixteen years earlier.  The apparent source followed the (large) proper motion of the nearby bright star over the intervening years, and there were null detections at the previous location in the current epoch, as well as at the future location in the earlier epoch, reducing the possibility of chance coincidences with unrelated high-energy objects.  The apparent X-ray brightness at Earth, averaged over the 98~ks of total exposure and accounting for absorption in the red giant's wind, is $\sim 2{\times}10^{-15}$ erg cm$^{-2}$ s$^{-1}$ (0.2--2~keV).  Systematic errors in the energy conversion factor, devolving from the unknown spectrum, amount to only about 10\%, smaller than the ${\sim}$30\% statistical uncertainties in the count rates.  The X-ray luminosity is only $3{\times}10^{25}$ erg s$^{-1}$, confirming Arcturus as one of {\em Chandra's}\/ darkest bright stars.
\end{abstract}


\section{INTRODUCTION}

Arcturus ($\alpha$~Bo\"otis; HD\,124897; K1.5~III)\footnote{Unless otherwise stated, stellar properties are from SIMBAD.} is the brightest star at northern declinations, third brightest overall.  It is an old, solar mass, slightly metal poor red giant, only 11~pc from the Sun (e.g., Ram{\'{\i}}rez \& Allende Prieto 2011).  Arcturus is of interest, among other reasons, because it mirrors the evolutionary fate that awaits the Sun some 5 billion years hence.  Another curiosity is that the red giant belongs to a well-defined stellar stream (the eponymous Moving Group): possibly the remnants of an ancient dissolved open cluster; more speculatively a tidal tail stripped from a satellite galaxy that wandered too close to the Milky Way eons ago (e.g., Navarro et al.\ 2004); or perhaps simply a dynamical resonance in the Galactic disk (e.g., Bensby et al.\ 2014).  

In the early days of X-ray astronomy, low-mass ($\gtrsim 1\,M_{\odot}$) red giants like Arcturus rarely were detected in high-energy surveys by pioneering observatories like {\em Einstein}\/ (Vaiana et al.\ 1981; Ayres et al.\ 1981), and later {\em R\"ontgensatellit} ({\em ROSAT}\,) (Haisch et al.\ 1991).  In contrast, yellow giants in the Hertzsprung gap, such as Capella ($\alpha$~Aurigae: G1~III + G9~III), often were strong coronal ($10^6$--$10^7$~K) emitters.  In fact, Linsky \& Haisch (1979) earlier had proposed -- based on ultraviolet proxies -- a dividing line in the giant branch, separating the coronal ``haves" from the ``have nots."  

The ``non-coronal'' side (redward of spectral type K1~III) later became known as the ``coronal graveyard,'' after a deep X-ray pointing by {\em ROSAT}\/ failed to detect Arcturus, prototype of the class (Ayres, Fleming, \& Schmitt 1991).  It was well known that an internal magnetic ``Dynamo'' -- relying heavily on stellar rotation  -- underpins the cycling activity of sunlike stars, so the demise of coronae among the bloated, slowly spinning red giants seemed sensible.  

A decade later, in mid-2002, one of the new-generation X-ray facilities, {\em Chandra,}\/  turned its sharper gaze on Arcturus (Ayres, Brown, \& Harper 2003 [ABH]).  In a 19~ks exposure with the High Resolution Camera (HRC-I), a mere 3 counts were recorded in a small detect cell centered at the predicted coordinates of the red giant.  Nevertheless, thanks to unusually low cosmic background conditions, the few counts represented a moderately significant detection.  The estimated X-ray luminosity of Arcturus was more than an order of magnitude lower than that of the average Sun, itself a rather mild coronal source.  The $L_{\rm X}$ was especially diminutive given that the surface area of the K giant is more than 600 times that of the G dwarf.

At about that time, the high-sensitivity ultraviolet spectrographs of {\em Hubble Space Telescope}\/ uncovered unexpected clues to the apparent coronal disappearing act.  The first surprise was the clear presence of coronal proxy \ion{C}{4} 1548~\AA\ in non-coronal giants like Arcturus (Ayres et al.\ 1997), albeit weak enough to have escaped previous notice.  \ion{C}{4} forms at $10^5$~K, hot enough that magnetic heating must be involved.  The second surprise was that other hot lines, \ion{Si}{4} 1393~\AA\  and \ion{N}{5} 1238~\AA, showed stationary, sharp absorptions from cool species such as \ion{Ni}{2} and \ion{C}{1} (ABH).  This implied that the hot emitting structures must be buried under a large overburden of lower temperature ($\sim 6000$~K) chromospheric material, a ``cool absorber'' if you will.  The large column can suppress soft X-rays, but still pass FUV radiation.  If the ``buried corona'' conjecture is correct, deep-seated magnetic activity on the non-coronal giants might be responsible for stirring up their atmospheres and initiating their powerful winds: $10^4$ times the Sun's mass loss rate (for the specific case of Arcturus), but much cooler than the solar coronal counterpart, $T\sim 10^4$~K versus $\sim 10^6$~K (e.g., O'Gorman et al.\ 2013).  The red giant outflows are important to galactic ecology, but the motive force behind the winds has remained elusive.

The best test of the buried corona hypothesis would be an X-ray spectrum, to judge the extent of the putative chromospheric soft absorption.  A minimal CCD-resolution ($E/\Delta{E}\sim 50$) energy distribution in the 0.25--10~keV band generally would require $\sim 10^3$ net counts; out of the question with contemporary instrumentation, at least given the apparent faintness of the Arcturus source in 2002.  However, recently the {\em Chandra}\/ Observatory offered a special opportunity to carry out observations that might help inform the design of next-generation X-ray facilities.  A proposal for a deeper HRC-I exposure of Arcturus -- in essence a feasibility assessment for a future spectrum -- was among the projects chosen.  Here, the results of the new X-ray observations of Arcturus are described, and their implications discussed.  To preview the more detailed conclusions presented later, a source at the predicted location of Arcturus (accounting for proper motion) was clearly present in each of the two new observations; and especially the sum (including also the earlier [2002] pointing).

\section{OBSERVATIONS}

{\em Chandra's}\/ HRC-I was the best choice for the project, because the sensor is immune to ``optical loading,'' an important consideration for observing visually bright stars that are X-ray faint.  HRC-I also has excellent low-energy sensitivity, important for low-activity coronal sources, which tend to be soft (taking the Sun as an example).  Further, {\em Chandra's}\/ high spatial resolution minimizes source confusion; and dilutes the diffuse cosmic background, as alluded earlier, which is essential to boost detectability of a source that might provide only a dozen counts in a long pointing.  The downside is that HRC-I has minimal spectral response:  it can deliver the  broad-band X-ray flux, but no clues to coronal temperature or soft absorption.  That limitation was not a serious concern, for what mainly was a detection experiment.

The new HRC-I observation of Arcturus was carried out in two segments, 50~ks and 30~ks on 2018 June 9 and 10, respectively.  Details are provided in Table~1, including the previous pointing from 2002, which was incorporated in the present analysis. 

\begin{deluxetable}{rrrrr}
\tabletypesize{\small}
\tablenum{1}
\tablecaption{{\em Chandra}\/ HRC-I Observations of Arcturus}
\tablecolumns{5}
\tablewidth{0pt}
\tablehead{\colhead{ObsID} & \colhead{UT Start} & \colhead{$t_{\rm exp}$} &  \colhead{${\Delta}x$} &  \colhead{${\Delta}y$} \\
\colhead{} & \colhead{} & \colhead{(ks)} & \colhead{(\arcsec)} & \colhead{(\arcsec)}  \\
\colhead{(1)} & \colhead{(2)} & \colhead{(3)} & \colhead{(4)} & \colhead{(5)}    
} 
\startdata
 2555       & 2002-06-19.02   &  18.39 &  $+0.3$ & $+0.3$ \\ 
20996      & 2018-06-09.40   &  49.43 &  $+0.2$ & $+0.1$ \\ 
21102      & 2018-06-10.82   &  29.77 &  $+0.4$ & $-0.1$  \\ 
\enddata
\tablecomments{Col.~3 is net exposure, corrected for dead time.  Cols.~4 and 5 are astrometric offsets determined from two X-ray sources with {\em Gaia}\/ optical counterparts.  Uncertainties in the offsets typically are $\sim{\pm}$0.1\arcsec.}
\end{deluxetable}

\section{ANALYSIS}

The three independent {\em Chandra}\/ pointings were considered by themselves, as well as together.  Fig.~1 depicts a time-integrated X-ray event map for the full 98~ks exposure: a 20\arcmin${\times}$20\arcmin\ field centered on coordinates (213.9120\arcdeg, $+$19.1766\arcdeg), approximately the mean position of Arcturus between the two separated epochs.  The event lists were concatenated in a fixed reference system, not accounting for the (large) proper motion of the target (which the other, more distant, objects in the vicinity were unlikely to share).  Perhaps three dozen X-ray point sources appear in the field, which is at high galactic latitude with a clear view out of the Milky Way.  Circled objects are from the {\em Chandra}\/ Source Catalog 2\footnote{see: http://cxc.harvard.edu/csc/}.  Those marked in red have {\em Gaia}\/ Data Release 2\footnote{see: https://gea.esac.esa.int/archive/} optical counterparts (most with high X-ray/optical ratios typical of Active Galactic Nuclei, although two of the barely detected objects apparently are distant late-type stars).  The three brightest X-ray sources with {\em Gaia}\/ counterparts -- likely all AGN -- were evaluated as checks of the aspect solution.  One of these -- {\em Gaia}\,1233978433822837888, to the upper right of the central region -- was slightly discrepant with respect to the other two (probably because of its vignetted profile owing to its large displacement from the image center) and was discarded.  Double circles mark the two remaining sources ({\em Gaia}\,1233964071455847296 and {\em Gaia}\,1233961631914413056) included in the final astrometric vetting.  Results are reported in Table~1: corrections were less than 0.5\arcsec, attesting to the excellent aspect reconstruction of {\em Chandra}\,.  A 20\arcsec${\times}$20\arcsec\ blow-up at upper left shows the central region around Arcturus in a map now accounting for the (large) proper motion of the nearby red giant.  A significant X-ray source is present at the co-moving optical position of the bright star.

For faint X-ray objects like Arcturus, the size of the detect cell -- to evaluate the number of source events -- must be chosen carefully.  Too large a cell accumulates more background, which can dilute the true source events and suppress the detection significance.  Too small a cell might throw away legitimate source counts, and also could be sensitive to subtle errors in the astrometric correction or knowledge of the encircled energy function (especially for an object of unknown spectral properties).  For the case of Arcturus, a 2\arcsec\ diameter detect cell ($\sim$90\% encircled energy) was adopted as a balance among these considerations.  

\clearpage
\begin{figure}[ht]
\figurenum{1}
\vskip  -30mm
\hskip  -5mm
\includegraphics[width=\linewidth]{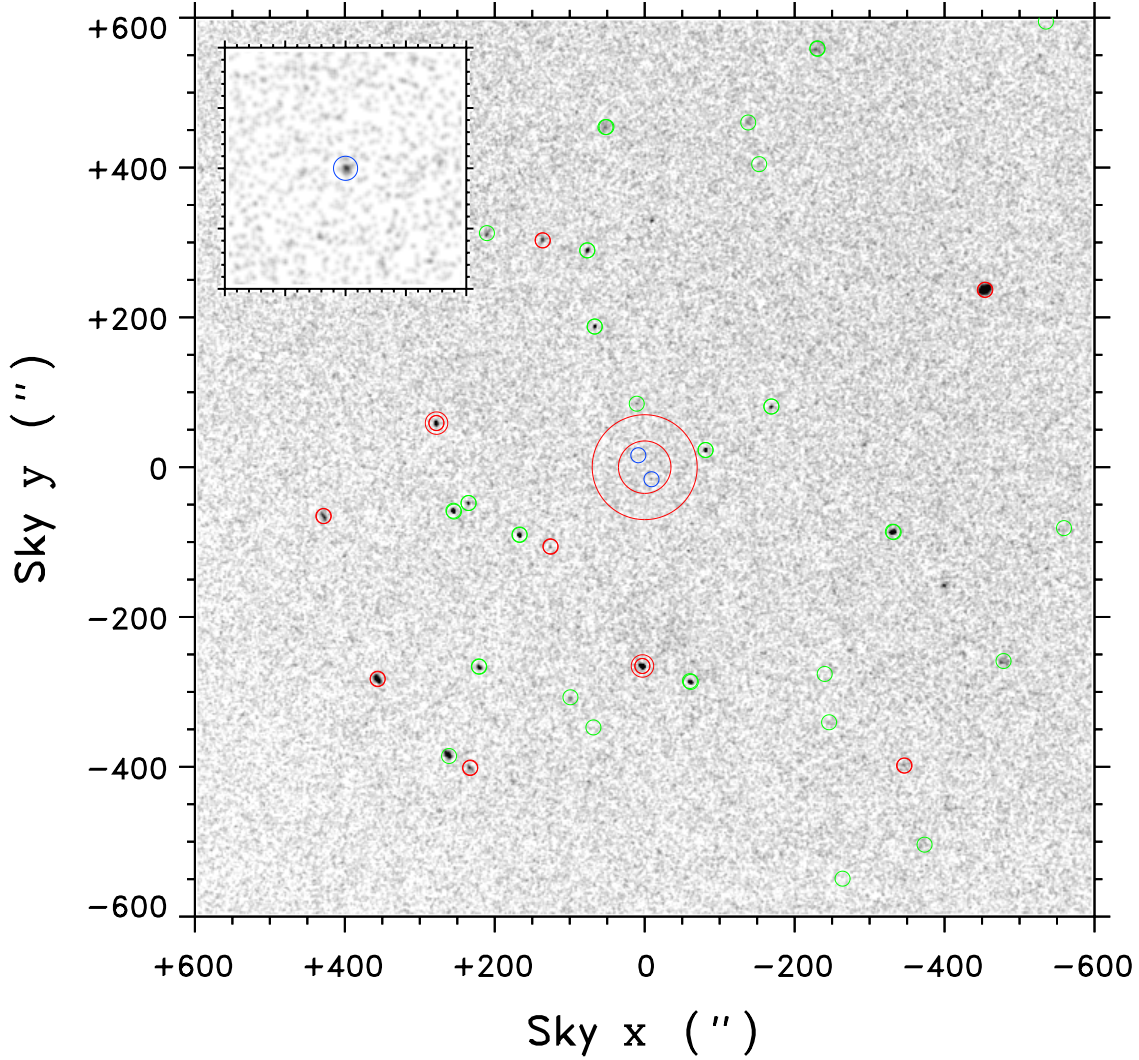} 
\vskip  -35mm
\caption[]{\small
Main panel displays central 20\arcmin${\times}$20\arcmin\ of {\em Chandra}\/ HRC-I field around Arcturus combining all three pointings (one in 2002, two in 2018), binned in 0.5\arcsec\ pixels and smoothed with a double-box-car spatial filter of 5~pixels width.  ``Sky'' coordinates are in arcseconds relative to a fixed reference point (213.9120\arcdeg, $+$19.1766\arcdeg).  N is up, E to the left.  Gray scale was set to highlight significant sources.  Pair of larger concentric red circles represents the annulus in which the diffuse cosmic background was assessed.  Two smaller blue circles, close to and on either side of center, mark positions of Arcturus in 2002 (upper left) and 2018 (lower right).  Additional small circles are entries from {\em Chandra}\/ Source Catalog 2.  Those in red have optical counterparts in {\em Gaia}\/ Data Release 2.  Double-circled objects (likely AGN) served as astrometric checks.  Inset panel at upper left is a 20\arcsec${\times}$20\arcsec\ field, binned in 0.125\arcsec\ pixels and smoothed with a double-box-car spatial filter of 3~pixels width, centered on the proper-motion corrected position of Arcturus.  Blue circle is 1\arcsec\ in radius: the detect cell for event measurements.  A highly significant source appears at the center of the cell.
}
\end{figure}

\clearpage
An average cosmic background was determined in a source-free annulus centered on the fixed reference coordinates in each epoch, as noted in Table~2 and illustrated in Fig.~1.  The background amounted to 4--6~counts in the detect cells of the more recent observations, but just 0.4 counts in the shorter, much lower background 2002 pointing.  Events not only were counted at the predicted location of the target in each epoch, but also in the 2002 observation at the coordinates where the target would be in the later 2018 pointing, and vice versa; to evaluate possible accidental sources.  The various measurements are summarized in Table~2.  Detection significances were based on the ``Frequentist" prescriptions described by Ayres (2004), while intrinsic source intensity confidence intervals were determined from the ``Bayesian'' prescriptions in the same article; reflecting the different statistical philosophies applied to source detection, on the one hand, and source characterization, on the other.

Note that the future location of Arcturus in the 2002 pointing, and its past location in the 2018 observations, have potential sources of much lower significance -- essentially null detections -- than the respective target cells in the same observations.  This suggests that the cumulative apparent source at Arcturus truly is the star, rather than accidental objects that happened to be at the precise stellar locations in the two well-separated epochs.  Notice also that the count rate (CR) of the second, shorter 2018 pointing was about 50\% higher than that of the first, which might suggest short term variability.  However, the two CR's agree within their 90\% confidence intervals, so variability cannot be claimed at a high level of significance. 

\begin{deluxetable}{lcccccc}
\tabletypesize{\small}
\tablenum{2}
\tablecaption{{\em Chandra}\/ HRC-I X-ray Measurements}
\tablecolumns{7}
\tablewidth{0pt}
\tablehead{ \colhead{Position} & \colhead{$N$} & \colhead{$B$} & 
\colhead{Detection $s$}  & \colhead{SL}  &  \colhead{$S_{0}$}  & 
\colhead{$[(S_{0})_{\Delta{S}-}^{\Delta{S}+}] / t_{exp}$} \\
\colhead{} & \colhead{(cnt)} & \colhead{(cnt)} & 
\colhead{}  & \colhead{(\%)}  &  \colhead{(cnt)}  & 
\colhead{(cnt ks$^{-1}$)} \\
\colhead{(1)} & \colhead{(2)} & \colhead{(3)} & \colhead{(4)} & \colhead{(5)}  & \colhead{(6)}   & \colhead{(7)}  
} 
\startdata
\cutinhead{ObsID\,2555: $t_{\rm exp}= 18.39$~ks; $b\sim 0.0073$ cnt ks$^{-1}$ $({\arcsec})^{-2}$ }
On Star  &     3    &  0.4  &  2.3  &   98.97  &   2.6   &   $0.14_{-0.07}^ {+0.20}$ \\[+2mm]
Future Position     &     0    &  0.4  &  0     &   \nodata &   0.0   &  $0.00_{-0.00}^ {+0.14}$ \\
\cutinhead{ObsID\,20996: $t_{\rm exp}= 49.43$~ks; $b\sim 0.0375$ cnt ks$^{-1}$ $({\arcsec})^{-2}$ }
On Star  &    17   &  5.8  &  3.6  &  99.98   & 11.2    &  $0.23_{-0.08}^ {+0.13}$ \\[+2mm]
Past Position     &     9    &  5.8  &  1.2  &  88       &   3.2     &   $0.06_{-0.05}^ {+0.10}$ \\
\cutinhead{ObsID\,21102: $t_{\rm exp}= 29.77$~ks; $b\sim 0.0376$ cnt ks$^{-1}$ $({\arcsec})^{-2}$ }
On Star  &   13    &  3.5  &  3.7  &  99.99  &   9.5     &   $0.32_{-0.12}^ {+0.20}$ \\[+2mm]
Past Position     &     5    &  3.5  &  0.7  &  76      &  1.5     &  $0.05_{-0.05}^ {+0.14}$ \\
\cutinhead{All: $t_{\rm exp}= 97.59$~ks; $b\sim 0.0318$ cnt ks$^{-1}$ $({\arcsec})^{-2}$ }
On Star  &    33   &  9.7  &  5.9  & 99.99  &   23.3    &   $0.24_{-0.06}^ {+0.09}$ \\[+2mm]
Future/Past Position     &    14   &  9.7  &  1.3  & 90       &    4.3     &   $0.04_{-0.03}^ {+0.06}$ \\
\enddata
\vskip -3mm
\tablecomments{In each heading, $b$ is the average background rate per unit area measured in the annulus $r=$~35\arcsec--70\arcsec\ centered on the fixed reference coordinates.  Col.~1 is notional location of the $r= 1\arcsec$ detect cell (90\% encircled energy): ``Future'' is the predicted 2018.5 position of the target in the 2002 pointing; ``Past'' is the 2002.5 position in the two 2018 observations.  Col.~2 is the total number of counts recorded in the cell.  Col.~3 is the expected number of background counts in the cell.  Col.~4 is the significance of the net source in one-sided Gaussian $\sigma$ (see: Ayres 2004).  Col.~5 is the significance level (SL) of the net source, in percent.  Col.~6 is net ``source'' counts in the cell, $S_{0}= (N - B) > 0$ (Bayesian prior is that the source cannot be negative, which is why the probabilities are one-sided).  Col.~7 is the derived count rate (CR, cnt ks$^{-1}$), where subscript and superscript specify a 90\% confidence interval (CI): ($S_{0} + \Delta{S}_{-}) \rightarrow (S_{0} + \Delta{S}_{+}$) (see: Ayres 2004), noting that $\Delta{S}_{-}$ is negative by convention.  The X-ray flux (0.2--2~keV) at Earth, $f_{\rm X}$ in erg cm$^{-2}$ s$^{-1}$, can be estimated by dividing CR values in Col.~7 by the encircled energy factor (0.9), then multiplying by the ECF ($\sim 8.1{\times}10^{-15}$ erg cm$^{-2}$ s$^{-1}$ (cnt ks$^{-1}$)$^{-1}$ for CR in cnt ks$^{-1}$.  The X-ray luminosity, in erg s$^{-1}$, can be obtained by multiplying $f_{\rm X}$ by the additional factor $4{\pi}d^2\sim 1.52{\times}10^{40}$ cm$^{2}$ for the $d= 11.26$~pc distance of Arcturus.}
\end{deluxetable}

While the count rate of Arcturus, now averaged over multiple epochs, is better established than the more tentative detection in the shorter 2002 pointing, an important -- potentially large -- uncertainty is the appropriate Energy Conversion Factor (ECF) to apply to the CR, especially lacking a spectrum to provide guidance concerning the source temperature and soft absorption.  It is helpful, in this regard, to temporarily ignore the main tenet of the ``buried corona'' conjecture, namely the possibly large internal soft X-ray absorption within the red giant chromosphere, because the ``un-absorbed flux'' corrections could be orders of magnitude, and thus essentially unconstrained at present (absent the desired future spectrum).  At the same time, it is important to consider the potential absorption effects of the extended red giant wind, outside the chromospheric attenuation zone, but between the star and the X-ray observatory at Earth. 

The O'Gorman et al.\ (2013) study, mentioned earlier, described a wind density model for Arcturus, consistent with centimetric free-free radio emission from the outflow, which implies a hydrogen density at the base of the wind (where $r\sim 1.2\,R_{\star}\sim 2.1{\times}10^{12}$~cm) of $\sim 3.8{\times}10^{7}$~cm$^{-3}$.  (By way of reference, for the stellar and wind parameters assumed by those authors, the mass loss rate would be a ${\rm few}{\times}10^{-10} M_{\odot}$~yr$^{-1}$.) For a homogeneous, radially expanding, constant velocity wind, the implied hydrogen column density through the outflow would be $\sim 8{\times}10^{19}$~cm$^{-2}$, much larger than the likely interstellar column in that direction to the nearby star.  The effect of the wind absorption is to flatten the ECF for the un-absorbed flux (i.e., the intensity if the outflow were not present) as a function of the source temperature (proxy for the spectral energy distribution).  

Over the broad temperature range $\log{T}\sim 6.4$--7.0~K, the average ECF is  $8.1{\pm}0.3{\times}10^{-15}$ erg cm$^{-2}$  s$^{-1}$ (cnt ks$^{-1}$)$^{-1}$, based on simulations with a solar abundance APEC model in WebbPIMMS\footnote{see: http://cxc.harvard.edu/toolkit/pimms.jsp}, with the cited wind column, to convert the measured CR (cnt ks$^{-1}$) to apparent (un-absorbed) X-ray flux (0.2--2~keV) at Earth.  For a softer spectrum, in the range $\log{T}\sim 6.0$--6.3~K, the ECF is only about 10\% higher; and for the ``absorbed'' flux (i.e., including the wind attenuation), the ECF is only about 10\% lower, at least for the warmer temperature interval.  The modeled ECF is insensitive to the assumed coronal abundances, over the values (0.2--1 solar) covered by WebPIMMS; and the detector sensitivity declined only slightly over the sixteen years between the two Arcturus observations, leading to a nearly negligible increase in the ECF.  

The apparent X-ray flux -- as measured at Earth, compensating for the wind absorption -- from the epoch-average CR is $f_{\rm X}\sim 2.2{\times}10^{-15}$ erg cm$^{-2}$ s$^{-1}$ (0.2--2~keV) with a formal uncertainty (from the CR alone) of about ${\pm}$30\%.  The systematic error on the un-absorbed flux, considering that the source might be softer -- $\log{T}\sim 6.0$~K compared with $\log{T}\sim 6.4$--7.0~K -- is only about 10\%.  The corresponding wind-free X-ray luminosity, for the 11.26~pc distance of the red giant, is $L_{\rm X}\sim 3.3{\times}10^{25}$ erg s$^{-1}$; while $L_{\rm X}/L_{\rm bol}\sim 5{\times}10^{-11}$.  The latter is a remarkable several orders of magnitude below the $\sim 1.5{\times}10^{-7}$ of the long-term average Sun, already close to the lowest activity tier among the G dwarfs of the solar neighborhood.  

The new flux values for Arcturus are about 2 times higher than those originally reported for the 2002 {\em Chandra}\/ pointing by ABH.  This partly is because the count rates in the new pair of 2018 pointings are about twice the earlier levels, but partly because a somewhat smaller ECF was assumed in the previous study.  The apparent CR up-tick in 2018 was welcome, given the 5 times higher background levels of those pointings (during solar minimum when external cosmic rays are more able to penetrate the inner heliosphere).  In any event, the very low measured X-ray luminosity of Arcturus in the total HRC-I observation still represents a stunning degree of coronal futility, although one should keep in mind that the apparent $L_{\rm X}$ could be the outcome of severe degradation by internal absorption in the extended red giant chromosphere.

\section{DISCUSSION}

Early speculations concerning the fading of X-ray activity in the coronal graveyard focused on the likely absence of a solar-like Dynamo in the evolved giants, which not only are rotating slowly, but also have a significantly different internal constitution compared with dwarf stars (a hydrogen burning shell around an inert helium core in first ascent giants; which gives way to core helium burning plus the hydrogen shell source in the post-flash objects, especially those in the long-lived red giant ``clump'').  However, it has become clear that classical Dynamos are only one part of the complex story of late-type stellar magnetism.  For example, much of the magnetic flux in the Sun's photosphere, especially at the minimum of sunspot activity, apparently is created in the near-surface layers and recycled rather quickly (days), populating what has been called the ``magnetic carpet'' (Title \& Schrijver 2002).  The generation mechanism likely is a purely convective process, without the necessity of rotation, so can operate in any star with surface convection, a condition the red giants certainly satisfy.  

In fact, Sennhauser \& Berdyugina (2011) reported a possible weak, $\lesssim 1$~G, longitudinal field on Arcturus; while, later, Auri{\`e}re et al.\ (2015) described similar detections on dozens of red giants, including Arcturus itself.  Although many of the red giants most similar to Arcturus in chromospheric activity displayed rather weak, sub-Gauss fields, it should be noted that the global longitudinal field of the Sun is only $\sim$2~G, which nevertheless is deceptively small because it represents an average over a sparse distribution of more intense, kilo-Gauss flux tubes.  

The key difference between a red giant and a yellow dwarf like the Sun would be the relative vertical scale of the surface magnetic structures, compared with, say, the thickness of the high-opacity outer atmosphere.  As described by ABH, the scale of the stellar outer convection zone, as a fraction of the star's radius, is similar for a giant and a dwarf.  Convectively spawned magnetic flux ropes likely would imprint at some characteristic fraction (say, a tenth) of that scale.  At the same time, the density scale height for a large diameter giant of similar mass to a small diameter dwarf (e.g., Arcturus and the Sun), will be a significantly larger fraction of the stellar radius, because the latter appears squared in the scale height relation (due to the gravity factor).  Because the density scale height controls the thickness of the stellar chromosphere, that layer will be proportionately much thicker in the red giant.

We know from the Sun that the ``relentlessly dynamic'' chromosphere (de~Pontieu et al.\ 2007) inspires an equally relentless forcing of the corona via hot plasma jets launched by magnetic reconnection in the kinematically stressed, tangled fields of the lower layers; especially in the narrow subduction lanes of the supergranulation pattern, where the magnetic carpet flux tubes accumulate, having been swept there by horizontal flows.  Perhaps a similar mechanism operates near the base of the Arcturus chromosphere, but affected by the great thickness of that region.  Most of the red giant magnetic loops might be buried well inside the chromosphere.  Those intercepting, and trapping, hot gas from reconnection jets at the base of the chromosphere would suffer X-ray attenuation.  However, any ballistic plasma jets threading onto open field lines in the chromosphere might burst free from that layer altogether, especially if there was an internal sustaining source of acceleration, for example MHD waves (as de~Pontieu and collaborators have proposed for solar spicules).  These renegade gas plumes, cooling rapidly by adiabatic expansion in the absence of confinement, might then become the source of the red giant wind.

Of course, all this is bare speculation without further quantitative insight concerning the organization of the outer atmospheres of red giants like Arcturus.  Such insight might come from a future advanced X-ray observatory, capable of delivering energy resolved spectra of these intrinsically faint coronal objects.  In the interim, alternative approaches could be pursued, such as further exploration of the cool absorptions on top of the red giant FUV hot lines; or possibly even orthogonal forays into other wavebands, such as the mm/sub-mm with ALMA.

One final note: Verhoelst et al.\ (2005) proposed -- from an analysis of infrared interferometric visibilities, and consistent with an earlier suggestion based on {\em Hipparcos}\/ -- that Arcturus might have a close companion, possibly a subgiant of mid-G spectral type.  The mass ratio would have to be very close to unity, with each component $\sim 1\,M_{\odot}$, to have two evolved stars in the same system.  Given that the proposed separation of the two nearly equal-mass stars is only 0.2\arcsec, the lack of periodic radial velocity variations is challenging to explain in the binary hypothesis, requiring a delicate tuning of the orbital configuration.  In fact, low amplitude radial velocity oscillations have been recorded in Arcturus (e.g., Merline 1996), but these appear to be stochastic, more closely related to solar pressure-modes than to systematic orbital effects.  The existence of a companion has not yet been confirmed by direct AO imaging at large telescopes, so remains in doubt.  More significantly, solar neighborhood late-type subgiants tend to have moderate X-ray luminosities ($\gtrsim 10^{28}$ erg s$^{-1}$: Schmitt \& Liefke 2004), so the apparent very low $L_{\rm X}$ of Arcturus would seem to further discount the binary hypothesis.

\section{CONCLUSIONS}

{\em Chandra}\/ pointings on the archetype non-coronal red giant Arcturus have secured moderately significant to very significant detections of an X-ray source at the stellar coordinates, in three epochs.  Accidental X-ray objects at the two distinct locations in 2002 and 2018 (well-separated thanks to high proper motion of the bright star) are unlikely, given the lack of significant sources at the future and past positions in the respective epochs.  Further, the high spatial resolution of {\em Chandra}\/ naturally minimizes source confusion.

Although the apparent Arcturus X-ray source is rather faint, it nevertheless suggests that a future high-energy observatory with $\sim$100 times the contemporary {\em Chandra}\/ sensitivity, and similar or better spatial resolution, could collect a diagnostically valuable spectrum in a reasonable exposure ($\sim$100~ks).  Such observations could help assess the properties of possibly buried coronae in the extended outer atmospheres of red giants, and perhaps also contribute to resolving the puzzle of their enigmatic winds. 

\acknowledgments
This work was supported by grant SP8-19001X from Smithsonian Astrophysical Observatory, based on observations from {\em Chandra}\/ X-ray Observatory, collected and processed at {\em Chandra}\/ X-ray Center, operated by SAO under NASA contract.  This study also made use of public databases hosted by {SIMBAD}, at {CDS}, Strasbourg, France; and Data Release 2 from ESA's {\em Gaia}\/ mission (\url{https://www.cosmos.esa.int/gaia}), processed by {\em Gaia}\/ Data Processing and Analysis Consortium (\url{https://www.cosmos.esa.int/web/gaia/dpac/consortium}), funded by national institutions participating in the {\em Gaia}\/ Multilateral Agreement.   


\end{document}